\documentclass{article}

\newcommand{\nn}{\nonumber\\}
\newcommand{\p}[1]{(\ref{#1})}
\newcommand{\cZ}{{\cal Z}}

\newcommand{\bz}{{\bar z}}

\newcommand{\cbZ}{\overline{\cal Z}}
\newcommand {\fc}{{1 + \alpha^2 z {\overline{z}}}}
\newcommand {\fac}{{(1 + \alpha^2 z \overline{z})}}

\newcommand{\bF}{{\overline F}}
\newcommand{\bQ}{{\overline Q}}

\newcommand{\bpsi}{{\bar\psi}}

\newcommand{\ba}{\begin{array}}
\newcommand{\ea}{\end{array}}
\newcommand{\be}{\begin{equation}}
\newcommand{\ee}{\end{equation}}
\newcommand{\bea}{\begin{eqnarray}}
\newcommand{\eea}{\end{eqnarray}}
\newcommand{\bi}{\begin{itemize}}
\newcommand{\ei}{\end{itemize}}

\newcommand {\bD}{\overline{D}}

\usepackage{amscd,amsmath,amssymb}

\begin{document}
\thispagestyle{empty}

\begin{center}
{\large\bf N=4 supersymmetric mechanics with nonlinear chiral supermultiplet}\\
\vspace{0.5cm} {\large S.~Bellucci,${}^{a}$
 A.~Beylin,${}^{b,c}$
 S.~Krivonos,${}^{b}$  A.~Nersessian, ${}^{d}$ and
E.~Orazi${}^{a,e}$}
\end{center}
{\it ${}^a$INFN-Laboratori Nazionali di Frascati, Via E. Fermi 40,
00044 Frascati, Italy}\\
{\it ${}^b$JINR, Bogoliubov Laboratory of Theoretical Physics,
141980 Dubna, Russia} \\ 
{\it ${}^c$Moscow Institute of Physics and Technology,
 Russia}\\
{\it ${}^d$\it Artsakh State University, Stepanakert, Nagorny Karabakh;\\
Yerevan State University and
  Yerevan Physics Institute,
Yerevan,
Armenia}\\
{\it ${}^e$Dipartimento di Fisica, Universit\`a di Roma Tor Vergata, Via della Ricerca Scientifica 1, 00133 Roma, Italy}\\
{\sl E-mails: bellucci, orazi@lnf.infn.it,$\; $ beylin,
krivonos@thsun1.jinr.ru,$\; $
  arnerses@yerphi.am}

\begin{abstract}
 We construct $N=4$ supersymmetric mechanics using the $N=4$
nonlinear chiral supermultiplet. The two bosonic degrees of
freedom of this supermultiplet parameterize the sphere $S^2$ and go into the bosonic
components of the standard chiral multiplet when the radius of the
sphere goes to infinity. We construct the most general action and
demonstrate that the nonlinearity of the supermultiplet results in
the deformation of the connection, which couples the fermionic degrees of freedom with the background,
and of the bosonic potential. Also a non-zero magnetic field could appear in the system.
\end{abstract}
\subsubsection*{Introduction.}
Supersymmetric mechanics with $N\geq 4$ supersymmetry has plenty of features which make
it interesting, not only
due to its relation with higher dimensional theories, but also as an independent theory. For instance,
$N=4$ superconformal
symmetry in $d=1$ is related with the one-parameter supergroup $D(2,1;\alpha)$ \cite{vp}, while in
higher dimensions the superconformal group does not contain any parameters. Another
interesting feature is the diversity of off-shell multiplets of $N\geq 4$ supersymmetry in $d=1$.
For example, for $N=4, d=1$ supersymmetry there are five off-shell linear
finite supermultiplets \cite{GR}
and two {\it nonlinear} ones \cite{ikl1}.
 One of them is a  one-dimensional analog of
the $N=2, d=4$ nonlinear multiplet \cite{n2d4nlin}.
The second one (called in \cite{ikl1} nonlinear chiral supermultiplet - NCS)
 seems to have  no known higher-dimensional analogs. It  includes, as a limiting case,
 the standard chiral supermultiplet  and has the same component content as the latter.

The idea of the nonlinear chiral supermultiplets comes about as follows.
If the two bosonic superfields $\cZ$ and $\cbZ$ parameterize the
two dimensional sphere $SU(2)/U(1)$ instead of flat space, then
they transform under $SU(2)/U(1)$ generators with the parameters
$a, \bar a$ as \be\label{eq1} \delta \cZ = a +{\bar a} \cZ{}^2
,\quad \delta \cbZ = {\bar a} +a \cbZ{}^2, \ee With respect to the
same group $SU(2)$, the $N=4$ covariant derivatives could form a
doublet\footnote{
In the trivial situation,
when the covariant derivatives do not transform under $SU(2)$,
this $SU(2)$ has nothing to do with the R-symmetry of the
fermionic sector of the theory. In this case nothing interesting is
happening.}
 \be\label{eq2} \delta D_i = - a \bD_i ,\quad \delta
\bD_i={\bar a} D_i \;. \ee
Here, the covariant spinor derivatives $D^i,
\bD_j$ are defined in the superspace $\mathbb{R}^{(1|4)}$ by
\be
D^{i}=\frac{\partial}{\partial\theta_{i}}+i\bar\theta{}^{i}\partial_t\,,\;
\bD_{i}=\frac{\partial}{\partial\bar\theta{}^{i}}+i\theta_{i}\partial_t\,,\quad
\{D^i, \bD_j\} = 2i \delta^i_j \partial_t .
\ee
One may immediately check that the
ordinary chirality conditions $ D_i \cZ =0 ,\quad \bD_i \cbZ=0 $
are not invariant with respect to \p{eq1}, \p{eq2} and they should
be replaced, if we wish to keep $SU(2)$ symmetry. It is rather
easy to guess the proper $SU(2)$ invariant constraints
\be\label{eq4} D_i \cZ = - \alpha\cZ \bD_i \cZ ,\quad \bD_i \cbZ =
\alpha\cbZ D_i \cbZ \;,\qquad \alpha={\rm const}.\ee So, using the
constraints \p{eq4}, we restore the $SU(2)$
invariance, but the price for this is just the nonlinearity of the
constraints. Let us stress that $N=4, d=1$ supersymmetry is
the minimal one where the constraints \p{eq4} may appear,
because the covariant derivatives (and the supercharges) form a
doublet of $SU(2)$ which cannot be real.

In this paper we construct $N=4$ supersymmetric mechanics with two
bosonic   and four fermionic degrees of freedom
 starting
from the most general action for NCS in $N=4$ superspace (a
preliminary step in the construction of such system
was done in \cite{ikl1}). We find that the configuration space of the system
is defined by a connection different from the symmetric connection of the base space,
in contrast with standard $N=4$ sypersymmetric mechanics with linear
chiral supermultiplet. Also,  the superpotential terms give rise
to an interaction with the magnetic field which preserves $N=4$
supersymmetry. The potential term is also modified as compared to
$N=4$ mechanics with linear chiral supermutiplet. In a special
limit, our action admits a reduction to the well known case of $N=4$
mechanics (see, e.g.  \cite{{BP},{BN}}).

\subsubsection*{Superfields formulation.}
As we already mentioned, the  $N=4, d=1$ NCS involves one complex
scalar bosonic superfield $\cZ$ obeying the constraints
(\ref{eq4}). If the real parameter $\alpha \neq 0$, it is always
possible to pass to $\alpha=1$ by a redefinition of the
superfields $\cZ, \cbZ$. So, it has only two essential values
$\alpha=1$ and $\alpha=0$. The latter case corresponds to the standard
$N=4, d=1$ chiral supermultiplet.
Now one can write the most general $N=4$ supersymmetric Lagrangian in $N=4$ superspace\footnote{We use
the convention $\int d^2 \theta = \frac {1}{4} D^i D_i$, $\int d^2 \bar\theta = \frac {1}{4} \bD_i \bD^i$,
$\int d^2 \theta d^2 \bar\theta = \frac {1}{16} D^i D_i \bD_i \bD^i$.}
\be\label{action1}
S = \int\! dt d^2\theta d^2 \bar\theta\; K(\cZ,\cbZ)
+ \int\! dt d^2 \bar\theta\; F(\cZ) + \int\! dt
    d^2\theta\; \bF (\cbZ) \;.
\ee
Here $K(\cZ,\cbZ)$ is an arbitrary function of the superfields $\cZ$ and $\cbZ$, while $F (\cZ)$ and
$\bF(\cbZ)$ are arbitrary holomorphic functions depending only on $\cZ$ and $\cbZ$, respectively.
 Let us
stress that our superfields $\cZ$ and $\cbZ$ obey the nonlinear
variant of chirality conditions \p{eq4}, but nevertheless the last
two terms in the action $S$ \p{action1} are still invariant with
respect to the full $N=4$ supersymmetry. Indeed, the supersymmetry
transformations of the integrand of, for example, the second term in
\p{action1} read \be\label{dok1} \delta F(\cZ) = -\epsilon^i D_i
F(\cZ)+2i\epsilon^i \bar\theta_i\dot{F}(\cZ)-\bar\epsilon_i
{\overline D}{}^i F(\cZ) +2i\bar\epsilon_i\theta^i\dot{F}(\cZ) \;.
\ee Using the constraints \p{eq4} the first term in the r.h.s. of
\p{dok1} may be rewritten as \be\label{dok2} -\epsilon^i D_i F
=-\epsilon^i F_{\cZ} D_i \cZ= \alpha \epsilon^i F_{\cZ} \cZ \bD_i
\cZ \equiv \alpha \epsilon^i \bD_i \int d\cZ \; F_{\cZ} \cZ \;.
\ee Thus, all terms in \p{dok1} are either full time derivatives or
disappear after integration over $d^2 \bar\theta$.

The irreducible component content of $\cZ$, implied by \p{eq4}, does not depend on $\alpha$ and can
be defined as
\be\label{comp}
    z = \cZ| , \; \bz = \cbZ|, \; A = -i \bD{}^i \bD_i \cZ| , \;
    \bar{A} =-i D^i D_i \cbZ| , \; \psi^i = \bD{}^i \cZ| , \; \bpsi{}^i = -D^i \cbZ|,
\ee where $|$ means restricting expressions to $\theta_i=\bar\theta{}^j=0$.
All higher-dimensional components are expressed as time
derivatives of the irreducible ones. Thus, the $N=4$ superfield
$\cZ$ constrained by \p{eq4} has the same field content as the
linear chiral supermultiplet.

Due to the nonlinearity of the basic constraints \p{eq4}, the transformation properties of the components \p{comp}
also contain nonlinear terms\footnote{All implicit summations go from ``up-left'' to ``down-right'', e.g.,
$\psi\bpsi \equiv \psi^i\bpsi_i$, $\psi^2 \equiv \psi^i\psi_i$,  etc.}
\bea
&&\delta z = (\alpha\epsilon_i z + \bar\epsilon_i) \psi^i ,\quad
\delta \psi^i = \frac{i}{2}(\bar\epsilon^i + \alpha\epsilon^i z) A - 2i \epsilon^i \dot{z}
    + \frac{1}{2} \alpha\epsilon^i (\psi)^2,\quad \delta A = - 4 \epsilon^i \dot{\psi_i}, \nn
&&\delta \bz = (\alpha\bar\epsilon^i \bz - \epsilon^i) \bpsi_i,\quad
\delta \bpsi_i = \frac{i}{2}(\epsilon_i - \alpha\bar\epsilon_i \bz)\bar{A} + 2i \bar\epsilon_i \dot{\bz}
    - \frac{1}{2} \alpha\bar\epsilon_i (\bpsi)^2,\quad
    \delta \bar{A} = 4 \bar\epsilon^i \dot{\bpsi}_i . \label{tr1}
\eea
\subsubsection*{Components formulation.}
After integrating in \p{action1} over the Grassmann variables and eliminating the auxiliary fields $A, \bar{A}$ by
their equations of motion, we get the action in terms of physical components
\bea\label{action2}
S &=& \int dt \Biggl \{ g
    \dot{z} \dot{\bz} - i\alpha\frac{ \dot{z} \bz
    }{\fc}F_{z} + i\alpha\frac{\dot{\bz} z
    }{\fc}\bF_{\bz} - \frac{ F_{z} \bF_{\bz} }
    {g\fac^2} + \nn
&&     \frac{i}{4} \fac g \left[
     \psi^i \dot{\bpsi}_i - \dot{\psi}^i \bpsi_i
      + \psi^i \bpsi_i  \left ( \frac{g_{\bz}}{g} \dot{\bz}
        - \frac{g_{z}}{g} \dot{z} +
        \alpha^2  \frac{ z \dot{\bz} - \dot{z} \bz}{\fac} \right)+
        \alpha\frac{ \dot{\bz} \psi^2 + \dot{z} \bpsi{}^2}{\fac}
            \right ] +\nn
    && \frac{1}{4}(\psi)^2 \Bigl [
        \frac{2\alpha^2 \bz F_{z}}{\fac} -  F_{z z}
        +  F_{z} \frac{g_{z}}{g}
    \Bigr ] - \frac{1}{4}(\bpsi)^2 \Bigl [
                 \frac{2\alpha^2 z \bF_{\bz}}{\fac} -  \bF_{\bz \bz}
        +  \bF_{\bz} \frac{g_{\bz}}{g}
    \Bigr ] -\nn
    && \frac{1}{16} (\psi)^2 (\bpsi)^2 \Bigl [
        2\alpha^2 g + \fac^2 g_{z \bz }
        - \fac^2 \frac{g_{z } g_{\bz}}
        {g}
    \Bigr ]
    \Biggr \},
\eea
where
\be\label{def1}
 g(z,\bz)=\frac{\partial^2K(z,\bz)}{\partial z\partial\bz},\quad
F_z=\frac{dF(z)}{dz},\quad \bF_\bz=\frac{d\bF (\bz )}{d\bz} \;.
\ee
Using the Noether theorem one can find classical expressions for the conserved supercharges corresponding
to the supersymmetry transformations \p{tr1}
\bea\label{scharges}
&& Q^i =  g\dot{\bar z} \psi^i - \alpha g\dot{z} \bz \bpsi^i  -  \frac{i}{4}\alpha^2 g z (\psi)^2 \bpsi{}^i
    +  \frac{i}{4}\alpha g (\bpsi)^2 \psi^i
    - i \frac{\alpha \bz F_{z}}{\fc} \psi^i
    + i \frac{\bF_{\bz}}{\fc} \bpsi{}^i \;, \nn
&& \bQ_i = g\dot{z} \bpsi_i + \alpha g\dot{\bar z} z \psi_i +   \frac{i}{4}\alpha^2 z g (\bpsi)^2 \psi_i
    +  \frac{i}{4} \alpha g(\psi)^2 \bpsi_i
    + i \frac{\alpha z \bF_{\bz}}{\fc} \bpsi_i
    + i \frac{F_{z}}{\fc} \psi_i  \; .
\eea

{}From the  bosonic part of the action, given by the first line in (\ref{action2}),
one may conclude that  the system contains a {\it nonzero magnetic field} with  the potential
\be
{\cal A}_{0}=i\alpha\frac{F_z \bz dz}{\fc} - i\alpha\frac{\bF_{\bz}zd{\bz}
    }{\fc}\;\;,\;
    d{\cal A}_{0}=i\alpha\frac{F_z+\bF_{\bz} }{(\fc)^2}dz\wedge d\bz.
\label{A0}\ee
The strength of this magnetic field is given by the
expression
\be B=\alpha\frac{(F_z+\bF_{\bz})}{(\fc)^2 g}\;. \ee
The bosonic potential is also modified
\be V(z, \bz)=
\frac{F_z\bF_{\bz}}{(\fc)^2 g}. \ee
One could
represent the fermionic part of  the kinetic term as follows:
\be {\cal S}_{KinF}=\frac{i}{4}\int dt
(\fc )g\left(\psi\frac{D\bpsi}{dt}-\bpsi\frac{D\psi}{dt}\right),
\label{action3} \ee where
 \be D\psi=d\psi+\Gamma\psi dz+T^+\bpsi dz,\quad D\bpsi=d\bpsi+{\bar\Gamma}\bpsi d\bz +T^-\psi d\bz,
\ee
and
\be\label{tor}
\Gamma=\partial_z\log \left((\fc){g}\right), \quad T^\pm=\pm\frac{\alpha}{\fc}.
\ee
Clearly enough, $\Gamma$, $\bar\Gamma$, $T^\pm$ define the components of the connection defining the configuration
superspace. The components  $\Gamma$ and $\bar\Gamma$ could be identified with the components
of the symmetric connection on the base space equipped with the metric $(\fc )gdzd\bz$, while
 the rest does not have a similar interpretation.

Thus, we conclude that the main differences between the proposed $N=4$ supersymmetric mechanics with NCS and
the standard one is the coupling of the fermionic degrees of freedom to the background, via the deformed connection,
the possibility to introduce a magnetic field, and the deformation of the
bosonic potential.

\subsubsection*{Landau problem}
There is a distinguished case, when the underlying space is the sphere with $g=(\fc)^{-2}$.
Choosing $F'(z)=B_0/2+ic_0$, we get a free particle on the sphere
in a constant magnetic field of magnitude $B_0$, and a trivial (constant) potential
$V_0=(B_0/2)^2 +c^2_0$, i.e.  the Landau problem on the sphere. Notice that very recently the simplest supersymmetric
extention of the Landau problem was used for  developing the theory of supersymmetric quantum Hall effect \cite{Has}.
Our model gives the $N=4$ extended supersymmetric background for that theory.

For $F'(z)=\omega_0z$
we get the potential for the oscillator on $CP^1=S^2$ (which is an  exactly solvable system,  even
in the presence of a constant uniform magnetic field) \cite{BNY},
in the  non-uniform magnetic field $B=\alpha\omega_0(z+\bz)$.

One can observe that similar  systems on the Lobachevsky plane ($g=(1-z\bz)^{-2}$, $\alpha=1$)
behave very differently from the ones on the sphere.

With the previously choosen superpotential $F'(z)=B_0/2+ic_0$,
we get the potential of the Higgs oscillator on the Lobachevsky plane
(in the absence of a magnetic field
this system is not only exactly solvable, but also ``maximally integrable")
$V=\left((B_0/2)^2+c_0^2\right)\left(1+4z\bz(1+z\bz)^{-2}\right)$ \cite{higgs}
and the non-uniform magnetic field
$B=B_0\left(1+2z\bar z(1+z\bz)^{-2}\right)$.
It seems obvious that for $B_0=0$ the corresponding $N=4$ supersymmetric system
will be exactly solvable as well.
For  $F'(z)=\omega_0z$ the potential and the  magnetic field are given by the expressions
$V_0=\omega_0^2z\bz(1-4z\bz(1+z\bz)^{-2})$, $B=\omega(z+\bz)(1-4z\bz(1+z\bz)^{-2})$.

This asymmetry between the systems on the sphere and the Lobachevsky plane is not so surprising, if we
recall that the system was built by using the  chiral supermultiplet constructed on the coset
$S^2=SU(2)/U(1)$, rather than $SU(1,1)/U(1)$.

\subsubsection*{Conclusion}
In the present paper we constructed $N=4$ supersymmetric mechanics with the nonlinear chiral
supermultiplet. The main interesting peculiarities of constructed system are
the non-standard coupling of the fermionic sector to the background,
the appearance in the action of the interaction
with the magnetic field having the strength
$$
B=\alpha\frac{(F_z+\bF_{\bz})}{(\fc)^2 g},
$$
and the deformation of the
 bosonic potential
 $$ \frac{F_z\bF_{\bz}}{g}\to
 \frac{F_z\bF_{\bz}}{(\fc)^2 g}.
$$
Let us recall that usually the appearance of a magnetic field breaks supersymmetry,
although it preserves the form of the bosonic part of the potential. Here, we have the
opposite situation.
This allows us to include in the class of $N=4$ supersymmetrizeable systems
the Landau problem on the sphere and the Higgs oscillator
on the Lobachewski plane.

These results should be regarded as preparatory for more detailed study of supersymmetric mechanics
with nonlinear supermultiplets. One of the obvious  questions is the quantization of the system,
and the construction of its $2n$-dimensional generalization corresponding to the dependence of the Lagrangian
on $n$ nonlinear chiral multiplets. It is  interesting to get such a nonlinear analog
for $N=8$ supersymmetric mechanics (on special K\"ahler manifolds) \cite{BKN}
and for duality transformations of such a system.
Also,  it is  still unclear whether is it  possible to extend the
system  to higher space-time dimensions. Finally, let us recall
that in one dimension there is the possibility to turn the auxiliary bosonic variables into dynamical ones \cite{GR}.
Similarly, one may convert dynamical variables  into auxiliary ones. Until now, such a procedure has been applied only
to linear supermultiplets. It is  interesting to check whether  it works  for nonlinear
supermultiplets, as well.

\subsubsection*{Acknowledgements.}
S.K.  thanks A.~Shcherbakov for valuable discussions.
S.K. and A.N.  thank INFN -LNF for the warm
hospitality extended to them during the course of this work.
This research was partially supported by the European
Community's Marie Curie Research Training Network under contract
MRTN-CT-2004-005104 Forces Universe, as well as by INTAS grants 00-00254  and 00-00262,
RFBR-DFG grant No 02-02-04002, grant DFG No 436 RUS 113/669, and RFBR grant
No 03-02-17440.

\end{document}